\documentclass[namedreferences]{solarphysics}

\usepackage[hyperref,optionalrh
]{spr-sola-addons} 
\usepackage{graphicx}        
\usepackage{color}           
\usepackage{breakurl}        

\usepackage{bm} 
\usepackage{epstopdf}

\def\be{\begin{equation}}
\def\ee{\end{equation}}
\def\bea{\begin{eqnarray}}
\def\eea{\end{eqnarray}}
\def\nn{\nonumber}
\newcommand{\ms}{\noalign{\vspace{3pt plus2pt minus1pt}}}

\renewcommand{\vec}[1]{{\mathbfit #1}}

\newcommand{\uvec}[1]{ \hat{\mathbf #1} }

\newcommand{\grad}{ {\bf \nabla } }

\def\half{{\textstyle{1\over2}}}

\newfont{\myfont}{cmmib10}

\newcommand\bi{\bm}

\chardef\us=`\_

\begin{document}

\begin{article}

\begin{opening}

\title{Is Cyclotron Maser Emission in Solar Flares Driven by a Horseshoe Distribution?}

%
\author[addressref=aff1,corref,email={donald.melrose@sydney.edu.au}]{\inits{D.B.}\fnm{D.B.}~\lnm{Melrose}{\orcid{000-0002-6127-4545}}}
\author[addressref=aff1,email={michael.wheatland@sydney.edu.au}]{\inits{M.S.}\fnm{M.S.}~\lnm{Wheatland}{\orcid{000-0001-5100-2354}}}


%
\runningauthor{D.B. Melrose, M.S. Wheatland}
\runningtitle{Horseshoe Electron Distribution}

%

\begin{abstract}
Since the early 1980s, decimetric spike bursts have been attributed to electron cyclotron maser emission (ECME) by the electrons that produce hard X-ray bursts as they precipitate into the chromosphere in the impulsive phase of a solar flare. Spike bursts are regarded as analogous to the auroral kilometric radiation (AKR), which is associated with the precipitation of auroral electrons in a geomagnetic substorm. Originally, a loss-cone-driven version of ECME, developed for AKR, was applied to spike bursts, but it is now widely accepted that a different, horseshoe-driven, version of EMCE applies to AKR. We explore the implications of the assumption that horseshoe-driven ECME also applies to spike bursts. We develop a 1D model for the acceleration of the electrons by a parallel electric field, and show that under plausible assumptions it leads to a horseshoe distribution of electrons in a solar flare. A second requirement for horseshoe-driven ECME is an extremely low plasma density, referred to as a density cavity. We argue that a coronal density cavity should develop in association with a hard X-ray burst, and that such a density cavity can overcome a long-standing problem with the escape of ECME through the second-harmonic absorption layer. Both the horseshoe distribution and the associated coronal density cavity are highly localized, and could not be resolved in the statistically large number of local precipitation regions needed to explain a hard X-ray burst. The model highlights the ``number problem'' in the supply of the electrons needed to explain a hard X-ray burst.
\end{abstract}

%

\end{opening}

%
\section{Introduction}
\label{s:introduction} 

\begin{figure}    
   \centerline{\includegraphics[width=0.7\textwidth,clip=]{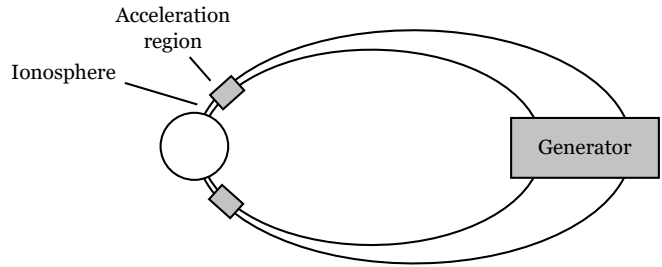}
              }
              \caption{Cartoon of a geomagnetic substorm showing the generator region, where electrons are assumed to be injected and field-aligned currents are generated, and the acceleration region, where the electrons that produce AKR through ECME are accelerated.}
   \label{Fig:generator}
   \end{figure}

Solar flares and magnetospheric substorms are both magnetic explosions, that is, they involve rapid release of magnetic energy that has been stored over a relatively long time. Despite notable differences between flares and substorms, such as the build up of magnetic energy being driven from below due to emerging magnetic flux  in a flare and from above due to the stress imposed by the solar wind in a substorm, there is a remarkable similarity in the phenomena observed. In both cases, a substantial fraction of the energy released goes into energetic electrons, that produce hard X-ray bursts (HXRBs) and type-III radio bursts in flares, and the visible aurora and auroral kilometric radiation (AKR) in substorms. The similarities have led to numerous exchanges of ideas between those modeling flares and substorms. A useful working hypothesis is that a phenomenon identified in one context has a counterpart in the other. One example is the transport of energy between the reconnection site and the acceleration region. In a substorm, the energy released by reconnection in the magnetotail is transported Alfv\'enically along field lines to the auroral acceleration regions, a few thousand kilometers above the ionosphere; a conventional model is illustrated schematically in Figure~\ref{Fig:generator}. A solar counterpart involves magnetic energy being released through reconnection relatively high in the corona, and transported Alfv\'enically along coronal magnetic field lines to an electron acceleration site near (or in) the chromosphere \citep{FH08,H12,MW13}. A related analogy involves the acceleration of electrons. There is compelling evidence for acceleration of auroral electrons by a parallel electric field $E_\parallel$, and such acceleration is the favored mechanism for the electrons that produce HXRBs \citep{H85}. In older literature, this acceleration was sometimes referred to a ``bulk energization'', reflecting the need to explain the extremely large number of electrons inferred to precipitate in a HXRB \citep{B71,B76}, in contrast with other acceleration processes in the solar corona \citep{Zetal11} that involve only a small fraction of the ambient plasma particles. 

\begin{figure}    
   \centerline{\includegraphics[width=0.7\textwidth,clip=]{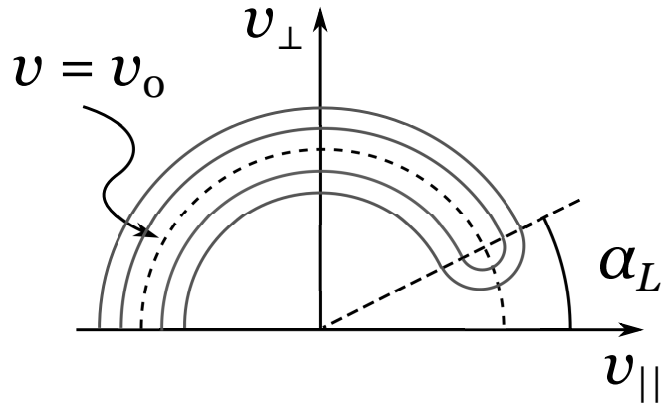}
              }
              \caption{Schematic of a horseshoe distribution in a 2D cross-section of velocity space, with the distribution azimuthally symmetric (independent of gyrophase). The dashed circle indicates the ring $v=v_0$ where the distribution function has its maximum, and the dashed line indicates the loss cone, $\alpha<\alpha_{\rm L}$, $\alpha=\arctan(v_\perp/v_\parallel)$. The solid curves indicated iso-contours of the distribution function.}
   \label{Fig:horseshoe}
   \end{figure}

In this paper we discuss another suggested analogy in detail: AKR is attributed to electron cyclotron maser emission (ECME) by the accelerated auroral electrons, and this is a counterpart to flare-associated solar spike bursts \citep{S78}, which are narrowband decimetric spikes that are strongly correlated with HXRs \citep{GAB91,Wetal11}. The analogy between spike bursts and AKR was first discussed in the 1980s when it was suggested that the ECME model also applies to flare stars \citep{HEK80,MD82a,SV84,DW87}. At that time, it was widely accepted that AKR is due to a loss-cone driven version of ECME \citep{WL79}. Subsequently, {\it in situ\/} data on the electrons that generate AKR showed them to have a horseshoe distribution \citep{Eetal00,BC00,Eetal02}. A horseshoe distribution, illustrated schematically in Figure~\ref{Fig:horseshoe}, may be defined as a combination of a ring distribution  (also called a shell distribution) and a loss-cone distribution. In a ring distribution particles are confined to a narrow range of speeds [$v$] such that the distribution function is sharply peaked, at $v=v_0$ say. A loss-cone distribution decreases sharply at pitch angles $\alpha<\alpha_{\rm L}$, where $\alpha_{\rm L}$ is the loss-cone angle.  A version of ECME driven by a horseshoe distribution is now widely accepted for AKR. An obvious question is whether horseshoe-driven ECME also applies to solar spike bursts, and to ECME from flare stars and other suggested astrophysical applications, such as shocks and blazar jets \citep{Betal03,Betal13}. A strong qualitative argument that this may be the case is based on the source of free energy that drives the ECME: in horseshoe-driven ECME the driver is the positive gradient in velocity space, $\partial f/\partial v>0$ at $v<v_0$. The free energy that drives this form of ECME is provided directly by the acceleration of the electrons by $E_\parallel$ in the direction of increasing magnetic field strength [$B$]. 

We address two specific questions in this paper. The first is:
Do the precipitating electrons in HXRBs have a horseshoe distribution?  The relevance of this question is that the evidence that the electrons that generate AKR have a horseshoe distribution is compelling, and the question arises as to whether this should be the case for precipitating electrons in other contexts. The arguments as to how a horseshoe distribution forms are qualitative \citep{T06} and need to be formalized before this question can be answered. The qualitative argument is based on three assumptions. First, electrons from near the apex of a flux loop are accelerated by $E_\parallel$ towards the ionosphere  (the chromosphere in the solar case). This increases the parallel component [$v_\parallel$] of their velocity. Second, the  magnetic moment $\mu=mv_\perp^2/2B$ of an electron is conserved as it propagates in the direction of increasing $B$. This partly redistributes the energy gained into the perpendicular component [$v_\perp$]. Third, some electrons mirror as they propagate towards the ionosphere, and those that would mirror below the ionosphere are lost. After mirroring the electron distribution has an upward-directed loss cone corresponding to the absence of the electrons that precipitated.  It needs to be shown how these assumptions imply a horseshoe distribution, that is, an isotropic ring distribution apart from a one-sided loss cone. 

The second question we address is:
Does horseshoe-driven ECME apply to solar spike bursts? There are two essential ingredients in the now-accepted version of ECME for AKR. One is that it is driven by a horseshoe distribution of energetic electrons accelerated by $E_\parallel$, and the other is that the source region is an ``auroral cavity'' which corresponds to a flux tube that is severely depleted of thermal plasma \citep{BC79,Aetal15}. Superficially, it seems that neither condition is satisfied in a solar flare. On the one hand, a horseshoe distribution is essentially mono-energetic, and observations of HXRBs are inconsistent with this, suggesting instead a power-law distribution \citep{B71,B76}. On the other hand, an extreme density depletion, like that in an auroral cavity (which is essentially devoid of thermal plasma) has not been proposed in the context of a solar flare, and so might seem implausible. Moreover, if the analogy applies, one might expect  spike bursts to be universally associated with HXRBs, as AKR is with auroral precipitation events, whereas only a few percent of HXRBs produce observable spike bursts \citep{GAB91}. In exploring the hypothesized analogy in more detail, we find counter-arguments against each of these objections, and no compelling reason to reject the hypothesis. Moreover, we also find that, if correct, the hypothesis has interesting wider implications concerning electron acceleration in flares.

In Section~\ref{sect:horseshoe} we present an analytic model that leads to a horseshoe distribution, and argue that it plausibly applies in a solar flare. In Section~\ref{sect:ECME} we discuss the various models for ECME, and point out that the free energy for horseshoe-driven ECME is provided by acceleration by $E_\parallel$. In Section~\ref{sect:discussion}, we discuss the relevance of horseshoe-driven ECME to solar spike bursts. The  conclusions are summarized in Section~\ref{sect:conclusions}.

\section{Formation of a Horseshoe Distribution}
\label{sect:horseshoe}

In this section we present a set of assumptions and show how they imply the formation of a horseshoe distribution. A qualitative description of the model we adopt is as follows.

The region where the electrons are ``injected'' is identified as the ``generator'' region, where the Alfv\'enic energy flux originates \citep{FH08,MW13}. We do not discuss the physics of the generator region in this paper, but we need to note several of its essential ingredients. First, this region acts as the energy sink for the magnetic energy stored and released in a large surrounding volume. Magnetic energy is transported into the generator region as a Poynting vector where it is partly converted into mechanical energy through magnetic reconnection. Second, the generator region acts as a source for the Alfv\'enic flux that transports the energy to an acceleration/dissipation region where the electrons are accelerated. The acceleration is due to $E_\parallel\ne0$ developing in an upward current region, where the current is carried by downgoing electrons. These electrons are assumed to be ``injected'' at the generator region, at the top of a magnetic flux tube.

\subsection{Specific Assumptions in the Model}

We are concerned with electrons that are accelerated in a solar flux tube such that they precipitate at the footpoints in the chromosphere, where they generate HXRBs through thick-target bremsstrahlung. We assume that the acceleration region is located well below the apex of the flux tube, and above the dense regions where the hard X-rays are emitted. 

The qualitative explanation given above of how a horseshoe distribution forms is based on downward acceleration by $E_\parallel$ being balanced by a redistribution into $v_\perp$ due to conservation of $\mu$. A potential [$\Phi_0$] is defined by  (minus) the integral of $E_\parallel$ along the field line. The loss cone forms due to downward propagating electrons with small $\sin\alpha$ being lost as they precipitate into the dense plasma. 

We argue that the following specific assumptions lead to a horseshoe distribution.
\begin{description}

\item[i)]  The electrons are ``injected''  at the apex of the flux tube with speed much less than $v_0=(2e\Phi_0/m)^{1/2}$ and are accelerated towards a footpoint by $E_\parallel$.

\item[ii)]  Pitch-angle scattering is negligible outside the injection region, so that the magnetic moment, $\mu=mv_\perp^2/2B$, is conserved.

\item[iii)] In a region around the ``injection'' point pitch-angle scattering is efficient, maintaining an isotropic electron distribution there.

\end{description}

 \begin{figure}    
   \centerline{\includegraphics[width=0.5\textwidth,clip=]{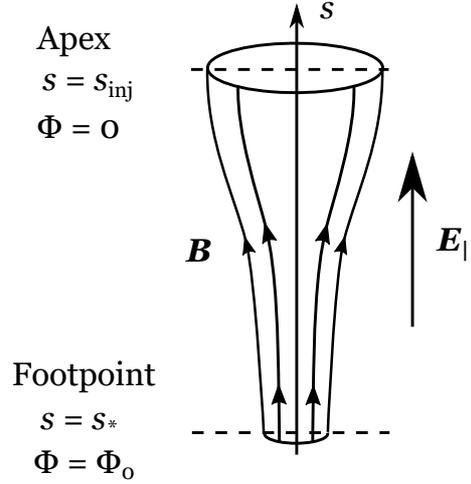}
              }
              \caption{The 1D model is illustrated as a vertical flux tube between the chromosphere (bottom) and the apex of the flux tube (top). Magnetic field lines are shown to diverge, as $B$ decreases from $s=s_*$ to $s=s_{\rm inj}$.}
   \label{Fig:geometry}
   \end{figure}

\subsection{1D Model}

Consider a one-dimensional (1D) model for electron motion in a flux loop. Let $s$ denote distance along a field line within the loop, increasing upward from one footpoint at $s=s_*$ to the injection point (at the apex of the flux tube) at $s=s_{\rm inj}$. The magnetic field at these two end points is $B_*=B(s_*)$ and $B_{\rm inj}=B(s_{\rm inj})$, respectively, with $B_*\gg B_{\rm inj}$. The 1D model has been used to determine the relation between the potential $\Phi(s)$ and the parallel current density [$J(s)$] in a self-consistent manner \citep{K73,W77,CS78,FL80}. Here the model, illustrated in Figure~\ref{Fig:geometry}, is used to discuss the formation of a horseshoe distribution. 

The downward acceleration of electrons by an upward electric field requires $E_\parallel(s)=-{\rm d}\Phi(s)/{\rm d}s>0$. The potential at the two points is $\Phi_*=\Phi(s_*)$ and $\Phi_{\rm inj}=\Phi(s_{\rm inj})$, respectively. The potential drop is $\Phi_0=\Phi_*-\Phi_{\rm inj}$, and we choose $\Phi_{\rm inj}=0$, $\Phi_*=\Phi_0$ without loss of generality. The potential energy of an electron [$-e\Phi(s)$] has its maximum ($=0$) at $s=s_{\rm inj}$, where the kinetic energy of an electron is assumed to be much smaller than $\half mv_0^2$. The kinetic energy increases as the potential energy decreases, and reaches its maximum above $s=s_*$, with this maximum corresponding to the initial kinetic energy plus $\half mv_0^2$.

\subsubsection{Two Constants of the Motion}

Along the field line between the apex and the footpoint, the motion of an electron is determined by two conserved quantities: its total energy [${\cal E}$] and its magnetic moment [$\mu$]. At any point $s$, the velocity components $v_\perp(s)=v(s)\sin\alpha(s)$ and $v_\parallel(s)=v(s)\cos\alpha(s)$ are determined by their value at the injection point and by the two constants. By solving
\be
{\cal E}=\half m[v_\perp^2(s)+v_\parallel^2(s)]-e\Phi(s)=\half mv_{\rm inj}^2,
\qquad
\mu={mv_\perp^2(s)\over2B(s)}= {mv_{\rm inj}^2\sin^2\alpha_{\rm inj}\over 2B_{\rm inj}},
\label{Hmu1}
\ee
with $\Phi_{\rm inj}=0$, one can determine $v_\perp,v_\parallel$ or $v,\alpha$ anywhere along the field line $s_{\rm inj}>s>s_*$. The solutions for $v_\perp,v_\parallel$ are
\be
v_\parallel(s)=\pm\left\{{2\over m}[{\cal E}-\mu B(s)+e\Phi(s)]\right\}^{1/2},
\qquad
v_\perp(s)= \left[{2\mu B(s)\over m}\right]^{1/2},
\label{Hmu2}
\ee
respectively, with the $\pm$-solutions interpreted as upgoing and downgoing electrons, respectively. The speed and pitch angle follow from Equation (\ref{Hmu2}), which implies
\be
v(s)=\left\{{2\over m}[{\cal E}+e\Phi(s)]\right\}^{1/2},
\label{v(s)}
\ee
and
\be
\cos\alpha(s)= \pm\left[{{\cal E}-\mu B(s)+e\Phi(s)\over{\cal E}+e\Phi(s)}\right]^{1/2},
\qquad
\sin\alpha(s)=\left[{\mu B(s)\over{\cal E}+e\Phi(s)}\right]^{1/2}.
\label{alpha(s)}
\ee
Electrons mirror at $v_\parallel(s)=0$ or equivalently $\sin\alpha(s)=1$, provided $s>s_*$. Electrons with $\sin\alpha(s_*)<1$ precipitate and are lost.

\subsubsection{Loss Cone}

The loss-cone angle is determined by considering an electron that mirrors at the footpoint. This corresponds to $v_\parallel(s_*)=0$, or $\sin\alpha(s_*)=1$, implying $\mu B_*={\cal E}+e\Phi_0$, which depends on both $\mu$ and ${\cal E}$. It is convenient to write this condition as $\mu=\mu_{\rm L}$, in terms of a limiting magnetic moment $\mu_{\rm L}$:
\be
\mu_{\rm L}={{\cal E}+e\Phi_0\over B_*}=\mu_{\rm L0}+{{\cal E}\over B_*}\approx\mu_{\rm L0},
\qquad
\mu_{\rm L0}={e\Phi_0\over B_*}={mv_0^2\over2B_*},
\label{muL}
\ee
where the approximation applies for ${\cal E}\ll\half mv_0^2$. Electrons with $\mu>\mu_{\rm L}$ mirror above the footpoint, and return to the apex of the flux tube. These upward directed electrons have a loss cone, corresponding the absence of electrons with $\mu<\mu_{\rm L}$. The loss-cone angle [$\alpha_{\rm L}(s)$] is found from Equation (\ref{muL}) and $\mu_{\rm L}=mv^2(s)\sin^2\alpha_{\rm L}(s)/2B(s)$. This gives
\be
\alpha_{\rm L}(s)=\arcsin\left[{B(s)\over B_*}\,{{\cal E}+e\Phi_0\over{\cal E}+e\Phi(s)}\right]^{1/2}
\approx\arcsin\left[{B(s)\over B_*}\,{\Phi_0\over\Phi(s)}\right]^{1/2},
\label{alphaL}
\ee
where the approximate expression applies for ${\cal E}\ll e\Phi(s)$.

A version of this model was solved to find $\Phi(s)$ and $J(s)$ \citep{K73,W77,CS78}. The solutions imply that $\Phi(s)$ is approximately proportional to $B(s)-B_{\rm inj}$, sometimes called the Knight relation. An implication of this approximate linear relation is that the loss-cone angle remains approximately constant as a function of $s$. 

\subsection{Electron Distribution Function}

The electron distribution function in this model can be inferred from arguments based on Liouville's theorem. An alternative derivation, based on the Boltzmann equation, is given in  Appendix~\ref{sect:Boltzmann}.

\subsubsection{Derivation Using Liouville's Theorem}

Liouville's theorem is that the the distribution function, in 6-dimensional ${\bi x}$-${\bi p}$ phase space, is a constant along the trajectory (in phase space) of a particle. The trajectory of the particle is found formally by solving the dynamical (Hamilton's) equations of motion, which we have effectively done in Equations (\ref{Hmu2})--(\ref{alpha(s)}). It is important to note that pitch-angle scattering must be excluded in order to appeal to Liouville's theorem: any form of diffusion invalidates the concept of a well-defined trajectory in phase space. 

Liouville's theorem becomes trivial if the distribution function can be written in terms of the constants of the motions, here ${\cal E}$ and $\mu$. The theorem then implies that the distribution is a set function, $F({\cal E},\mu)$, at all points $s$ along a trajectory. A minor complication is that due to the $\pm$ solutions in Equation (\ref{Hmu2}), one needs to introduce two separate functions [$F_\pm({\cal E},\mu)$]. Liouville's theorem implies 
\be
f(v_\perp,v_\parallel,s)=F_+({\cal E},\mu)+F_-({\cal E},\mu).
\label{Hmu3}
\ee
The functions $F_\pm$ become functions of the variables $v_\perp,v_\parallel,s$ by simply writing ${\cal E}$ and $\mu$ in terms of these variables, as is done in Equation (\ref{Hmu1}). The distribution function, $f_{\rm inj}$ say, is assumed to be given at $s=s_{\rm inj}$, and one writes $f_{\rm inj}$ as a function of ${\cal E}$ and $\mu$ at $s=s_{\rm inj}$. Specifically, one uses
\be
F_\pm({\cal E},\mu)=f\big((2\mu B/m)^{1/2},\pm[({\cal E}-\mu B+e\Phi)/m]^{1/2},s\big)
\label{Hmu3a}
\ee
to identify $F_\pm({\cal E},\mu)$ at $s=s_{\rm inj}$ using the given $f_{\rm inj}$, and then uses Equation (\ref{Hmu3}) to identify the distribution function at an arbitrary position $s$. 

It is helpful to make a further separation of $F_\pm$ by writing
\be
F_\pm({\cal E},\mu)=F^>_\pm({\cal E},\mu)+F^<_\pm({\cal E},\mu),
\label{Hmu3b}
\ee
where the superscripts indicate $\mu>\mu_{\rm L}$ and $\mu<\mu_{\rm L}$, respectively. Consider downgoing electrons. Those in the distribution $F^>_-$ mirror above the footpoint. After mirroring these form the distribution $F^>_+$ of electrons that return to the apex. Hence one has $F^>_+=F^>_-$. There are no upgoing electrons with $\mu<\mu_{\rm L}$, implying $F^<_+=0$.

\subsubsection{Isotropy Requirement}

Two notable features of the observed horseshoe distributions are that the distribution is (approximately) independent of pitch angle [$\alpha$] outside the loss cone, and that the loss cone is one sided. In this case, the contours of constant $f$ form an incomplete ring in velocity space with a gap at $\alpha<\alpha_{\rm L}$, as illustrated in Figure~\ref{Fig:horseshoe}. The constant [${\cal E}$] is independent of $\alpha$ but $\mu$ depends explicitly on $\alpha$. It follows that an isotropic distribution must be independent of $\mu$. 

For the electrons that generate AKR, the loss cone is one sided, implying that the electrons are not bouncing back and forth between mirror points in the two hemispheres, which would imply loss cones at both $\alpha<\alpha_{\rm L}$ and $\alpha>\pi-\alpha_{\rm L}$. It is essential that a model for the formation of a horseshoe distribution   in the magnetosphere be consistent with this one-sidedness. A plausible explanation is that the electrons are isotropized on returning to the injection region at the apex of the loop. This is the reason for our assumption (iii). In particular, the returning electrons, with distribution $F_+$, are anisotropic due to the loss cone, and if the electrons were not efficiently scattered at the looptop, this would result in downgoing electrons with a loss cone in the conjugate hemisphere. Pitch-angle scattering must be strong, in the sense that the electrons diffuse sufficiently rapidly to fill the loss cone before the returning electrons escape the looptop region again. In other words, an electron returning to the injection region from either hemisphere must lose memory of where it came from before escaping from the injection/generator region. 

In the solar case, such efficient pitch-angle scattering near the apex of the flux tube is one notable difference between the model proposed here and the model of \citet{EH95}. An implication is that the electrons drawn upward from one footpoint join a pool of efficiently scattered electrons in the generator region, and this pool is the source of the ``injected'' electrons in our 1D model. We do not model the generator region in detail in this paper, but we note that an essential feature of our model, compared with that of \citet{EH95}, is that all current loops close across field lines in the generator region and in the chromosphere at one footpoint. The cross-field current in the generator region is regarded as the driver (supplying the energy) to the Alfv\'enic flux that results in the electron acceleration near the footpoint. In contrast in a model with cross-field current closure at both footpoints \citep{EH95}, the driver must be a photospheric dynamo, and not a magnetic explosion.

\subsection{Maxwellian Model}

As an example,  which turns out to be too simple, suppose that in the injection region the electrons are isotropic with a Maxwellian distribution at a temperature $T_{\rm inj}$ (in energy units), and that the escaping electrons have the same distribution. This corresponds to a distribution,  inside the injection region where $\Phi$ is zero,
\be
f_{\rm inj}(v)={n\over(2\pi)^{3/2}V_{\rm inj}^3}\exp\left(-{v^2\over2V_{\rm inj}^2}\right),
\qquad
F_{\rm inj}({\cal E})={n\over(2\pi)^{3/2}V_{\rm inj}^3}\exp\left(-{{\cal E}\over T_{\rm inj}}\right),
\label{mm1}
\ee
with $T_{\rm inj}=mV_{\rm inj}^2$ and ${\cal E}=\half mv^2$. Outside the injection region one has $\Phi(s)\ne0$ and there is a loss cone at $\sin\alpha<\sin\alpha_{\rm L}(s)$ for upward directed electrons. The foregoing discussion implies that the distribution anywhere between the injection point and the footpoint can then be written as
\be
F({\cal E},\mu)=F_{\rm inj}({\cal E})H(\mu-\mu_{\rm L}),
\label{mm2}
\ee
where $H$ is the step function. Equation (\ref{mm2}) may be rewritten as
\be
f(v,s)={n\over(2\pi)^{3/2}V_{\rm inj}^3}\exp\left(-{\half mv^2-e\Phi(s)\over T_{\rm inj}}\right)H[\alpha-\pi+\alpha_{\rm L}(s)],
\label{mm3}
\ee
which is a Maxwellian form for a horseshoe distribution.

A Maxwellian distribution of the form (\ref{mm3}) has $\partial f(v)/\partial v<0$ and hence is unacceptable as a model for horseshoe driven ECME. The distribution of escaping electrons differs from the assumed Maxwellian in the source region due to faster particles escaping preferentially. A semi-quantitative model for this modification is to assume that the escaping distribution differs from that in the source region by a factor which is a power of $v_{\rm inj}$:
\be
F_{\rm esc}({\cal E})\propto{\cal E}^{b/2}\exp\left(-{{\cal E}\over T_{\rm inj}}\right).
\label{mm1}
\ee
The distribution (\ref{mm1}) describes a situation where the probability of escape is proportional to $v_{\rm inj}^b$. Such a distribution has $\partial f(v)/\partial v>0$ for ${\cal E}<bT_{\rm inj}/2$.

\subsection{Electron Current Density}

Various averages over the distribution function can be expressed in terms of $F_\pm$. To do this involves the integral over velocity space being rewritten in terms of integrals over ${\cal E}$ and $\mu$, as defined by Equation (\ref{Hmu1}). The Jacobian of the transformation is
\be
J\left\{\!
{{\cal E},\mu\;\atop
v_\parallel,v_\perp}
\!\right\}={m^2v_\parallel v_\perp\over B}.
\label{intHmu1}
\ee
For example, the number density of electrons is \citep{W77}
\bea
n&=&2\pi\int_{-\infty}^{\infty}{\rm d}v_\parallel\int_0^{\infty}{\rm d}v_\perp v_\perp\,f
\nn\\
\ms
&=&
{2\pi B\over m^2}\int_0^\infty {\rm d}\mu\int_{\mu B-e\Phi}^\infty {\rm d}{\cal E}\,{F_++F_-\over[2({\cal E}-\mu B+e\Phi)/m]^{1/2}}.
\label{intHmu2}
\eea
The only electrons that can carry this current are those described by $F_-^<$, that is, the electrons that propagate from the injection point and precipitate at the footpoint.
The parallel current density carried by the electrons is
\be
J_\parallel=
-{2\pi eB\over m^2}\int_0^\infty {\rm d}\mu\int_{\mu B-e\Phi}^\infty {\rm d}{\cal E}\,F_-^<({\cal E}),
\label{intHmu3}
\ee
which we write as $J(s)$ to emphasize its dependence on $s$.

Detailed models \citep{K73,W77,CS78,FL80} suggest that the current $J_\parallel\to J(s)$, given by Equation (\ref{intHmu3}), is approximately proportional to $B(s)$ over $s_{\rm inj}\ll s\ll s_*$. This corresponds to a total field-aligned current that is approximately constant (independent of $s$), except at the end points where it connects to cross-field currents. This applies to the upward-current region, which includes the acceleration region of the electrons. There is a downward field-aligned current on neighboring field lines, such that the current forms a circuit that closes across field lines in the generator regions and in the chromosphere.

\section{Drivers for ECME}
\label{sect:ECME}

The horseshoe model for ECME was originally motivated by observations of the auroral electrons that generate AKR. The significance of the model, and its relation to earlier models, are discussed in this section. It is relevant to start with a brief historical review of the development of the theory.

\subsection{Versions of ECME}

The earliest theory for ECME was proposed by \citet{T58}. Other early theories \citep{S59,BHB61} were similar to Twiss's theory in three notable ways. First, the emission was assumed to be at the relativistic gyrofrequency $\omega=\Omega_{\rm e}/\gamma\approx\Omega_{\rm e}(1-v^2/2c^2)$ where $\Omega_{\rm e}=eB/m$ is the cyclotron frequency. This implies a one-to-one correspondence between the frequency [$\omega$] and the speed [$v$] of the electron. Second, the driving term for the maser was assumed to be an inverted energy population, which in the isotropic case corresponds to $\partial f(v)/\partial v>0$ for $v<v_0$. An analytic model is a ring distribution of the form  $\propto(v^2/2V^2)^N \exp(-v^2/2V^2)$, which is sharply peaked around $v_0=(2N)^{1/2}V$ for large $N$ and is an isotropic counterpart of a DGH distribution \citep{DGH65}. Third, vacuum conditions were assumed, such that the refractive index is unity, $n=kc/\omega=1$. The emission was separated into two orthogonal linear polarizations, and although these were sometimes referred to as modes, they do not correspond to the relevant (o- and x-) modes of magnetoionic theory. 

Although these early theories for ECME were not motivated by any specific astrophysical application, at about the same time it was recognized that Jupiter's decametric radio emission (DAM) is emitted at the cyclotron frequency. An early cyclotron model for DAM \citep{E62}, which was based on emission by bunches, took the dispersive properties of the plasma into account. Cyclotron emission strongly favors the x-mode over the o-mode. A point emphasized by \citet{E62,E65} is that in order for ECME to escape, it needs to be Doppler shifted to above the cutoff frequency for the x-mode at
\be
\omega_{\rm x}=\half\Omega_{\rm e}+\half(\Omega_{\rm e}^2+4\omega_{\rm p}^2)^{1/2}\approx\Omega_{\rm e}+{\omega_{\rm p}^2/\Omega_{\rm e}},
\label{omegax}
\ee 
where the approximation applies for $\omega_{\rm p}\ll\Omega_{\rm e}$. The first ECME theory for DAM invoked a ring distribution and ignored the dispersive properties of the plasma \citep{HB63}. This version of ECME has $\omega<\Omega_{\rm e}$ and is inconsistent with the requirement $\omega>\omega_{\rm x}$ for escaping emission in the x-mode.

In the early 1970s, a terrestrial counterpart for DAM was identified \citep{G74}, and is now called AKR. \citet{M76} proposed a different version of ECME  for both DAM and AKR. This model, which includes the Doppler shift to $\omega>\omega_{\rm x}$, was based on the theory for cyclotron instability of (magnetoionic) waves in a plasma with an anisotropic electron distribution \citep{SS61}. However, this model requires an extreme form of anisotropy, as a driver, for which there was no evidence. \citet{WL79} proposed a model for ECME that includes both the relativistic correction to the cyclotron frequency and the Doppler shift to $>\omega_{\rm x}$. This version is driven by a  distribution with $\partial f/\partial v_\perp>0$, which condition is satisfied by a loss-cone distribution. This loss-cone-driven maser became the preferred version of ECME for well over a decade. It was applied not only to DAM and AKR, but also to solar spike bursts \citep{HEK80} and to radio emission from flare stars \citep{MD82a}.

The requirement $\omega_{\rm p}\ll\Omega_{\rm e}$, for loss-cone driven ECME to operate, was confirmed for AKR when it was discovered that the electron density, within the flux tube to which the inverted-V electrons are confined, is orders of magnitude lower than in the surroundings \citep{BC79}. Such a local, low-density region is referred to as an auroral cavity. It was assumed  \citep{MD82a} that the requirement $\omega_{\rm p}\ll\Omega_{\rm e}$ can be satisfied in the lower corona; this point was discussed in detail recently by \citet{R15}.

\subsection{Resonance Ellipse}
\label{sect:ellipse}

A useful concept in discussing ECME is a graphical interpretation of the gyro\-resonance condition
\be
\omega-s\Omega_{\rm e}/\gamma-k_\parallel v_\parallel=0,
\label{gres}
\ee 
which is the condition for an electron with given $v_\perp,v_\parallel$ to resonate with a wave with given $\omega,k_\parallel$ at the $s$th harmonic. The parallel wavenumber $k_\parallel=n(\omega/c)\cos\theta$ depends on the refractive index [$n$] of the wave mode, here assumed to be the x-mode. When plotted in $v_\perp$-$v_\parallel$ space for given $\omega,k_\parallel,s$, Equation (\ref{gres}) defines a resonance ellipse \citep{OG82,MRH82,M86}. The ellipse (actually a semi-ellipse with the region $v_\perp<0$ unphysical) is centered on the $v_\parallel$-axis, at a point $\propto k_\parallel$, with its major axis along the $v_\perp$-axis. The physical significance of the ellipse is that the absorption coefficient, which must be negative for ECME to occur, can be written as a line-integral around the ellipse. For a given distribution function, this allows one to identify the most favorable ellipse as the one that maximizes the negative contribution to the absorption coefficient. For cases of relevance here, the dominant driving term is $\propto\partial f/\partial v_\perp>0$, and the largest growth rate corresponds to the ellipse that maximizes the (weighted) contribution from this term.

\begin{figure}    
   \centerline{\includegraphics[width=0.5\textwidth,clip=]{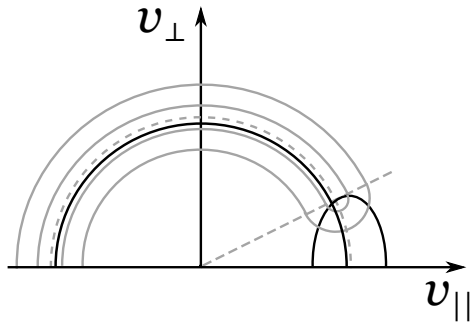}
              }
              \caption{Two resonance ellipses are illustrated for a horseshoe distribution. One is circular, corresponding to perpendicular emission, through the region just below $v=v_0$ where the ring distribution has its maximum positive value of $\partial f/\partial v$. The other indicates the ellipse that gives the maximum contribution from $\partial f/\partial v_\perp$ associated with the loss-cone feature, corresponding to loss-cone driven ECME.}
   \label{Fig:ellipse}
   \end{figure}

The maximum growth rate for a ring distribution, at $v=v_0$, is for the case where the ellipse reduces to a circle centered on the origin, which corresponds to $k_\parallel=0$, as illustrated in Figure~\ref{Fig:ellipse}. The maximum contribution from $\partial f/\partial v>0$ corresponds to a resonance circle with a radius slightly less than $v_0$. It follows that the most favorable case for a ring distribution is ECME perpendicular to the field lines, $\theta=\pi/2$, at $\omega\approx\Omega_{\rm e}(1-v_0^2/2c^2)$.  The line-integral around the resonance circle is then an integral over pitch angle, with all pitch angles contributing. 

For a loss-cone distribution, the resonance ellipse that leads to the maximum growth rate depends on the distribution in $v$, as well as on the loss-cone distribution. For simplicity, consider the loss-cone feature in a horseshoe distribution, in which case the distribution in $v$ is strongly concentrated around $v=v_0$. The ellipse that gives maximum growth, due to $\partial f/\partial v_\perp>0$ inside the loss cone, has its center on the $v_\parallel$-axis, at $v_\parallel=k_\parallel c^2/\Omega_{\rm e}$ with $v_\parallel=v_0\cos\alpha_{\rm L}$, and a semi-major axis corresponding to $v_\perp=(\omega-\Omega_{\rm e})/k_\parallel$ with $v_\perp\approx v_0\sin\alpha_{\rm L}$. This resonance ellipse corresponds to a Doppler-shifted frequency $\omega\approx\Omega_{\rm e}(1-v_0^2/2c^2)+k_\parallel v_\parallel$, with $k_\parallel v_\parallel>0$. The magnitude of the growth rate is sensitive to the details of the ellipse, resulting in it being a sensitive function of $\omega-\Omega_{\rm e}$ and $\theta$. As a consequence, one expects loss-cone driven ECME to be confined to small ranges of $\omega$ and $\theta$ about the values that maximize the growth rate. The only electrons that contribute energetically to the ECME are those that lie on this ellipse and have $v\lesssim v_0$ and $\alpha\lesssim\alpha_{\rm L}$. There are many fewer electrons that contribute to the integral around this resonance ellipse than those that contribute to the integral around the resonance circle. As a consequence, the growth rate for loss-cone driven ECME is intrinsically much smaller than for ring-driven ECME.

In the case of a ring distribution the free energy that drives the maser is an energy inversion: $\partial f/\partial v>0$ at $v<v_0$ implies that there are more electrons with higher energy than with lower energy for $\half mv^2<\half m v_0^2$. These electrons lose energy to the waves. This loss of energy opposes the acceleration by $E_\parallel$ that causes the ring distribution to develop. These ingredients provide a simple model for the flow of energy from the accelerating electric field to the escaping ECME. The flow of free energy is less obvious for loss-cone driven ECME. The loss-cone angle [$\alpha_{\rm L}$] decreases with increasing height. The electrons that drive this form of ECME are those just inside the loss cone, and at any height these are the electrons whose $v_\perp$ implies the lowest possible mirror point.  The inverted spectrum is due to the removal, through precipitation, of electrons with $v_\perp$ lower than this value. The back reaction of the ECME is to drive electrons to lower $v_\perp$, tending to fill the loss cone at smaller $v_\perp$.

\subsection{Horseshoe-Driven ECME}

The maximum growth rate for horseshoe-driven ECME is similar to that for a ring distribution. As explained in Section~\ref{sect:ellipse}, the most favorable ellipse is a circle for a ring distribution, corresponding to perpendicular emission, $\theta=\pi/2$ at a frequency $\omega<\Omega_{\rm e}$, and the maximum growth for horseshoe-driven ECME due to the ring-type feature is approximately the same, with $\theta\approx\pi/2$ and $\omega\approx\Omega_{\rm e}(1-v_0^2/2c^2)<\Omega_{\rm e}$. The contribution to the growth rate from the loss-cone feature is smaller, and is usually neglected. (A rough estimate of the ratio of the growth rates is the ratio of the lengths of the arcs of the resonance ellipses through the region where the distribution function has its maximum.) However, it is relevant to note that ring-driven and loss-cone-driven ECME lead to growth of waves at different frequencies and angles, and one can imagine conditions where the slower-growing loss-cone-driven ECME could be observable. We do not discuss this point further here.

In the application of horseshoe-driven ECME to AKR, the requirement $\omega>\omega_{\rm x}$ is assumed not to be relevant. The argument is that when $\omega_{\rm p}$ is sufficiently low (and the plasma is sufficiently hot) vacuum-like wave dispersion applies. Specifically, the stop band between magnetoionic z~and x-modes is assumed to be washed out, so that emission below the cyclotron frequency can escape. The observational evidence is that the source region (auroral cavity) is essentially devoid of thermal plasma and that vacuum-like dispersion is indeed a valid approximation \citep{Petal02,Setal14}. The emitted radiation is assumed to be ducted upward, by reflection from the cavity walls, until it reaches a height where it can escape. Escape becomes possible at a height where the cutoff frequency [$\omega_{\rm x}$] of the x-mode outside the flux tube is below the wave frequency.

\begin{figure}    
   \centerline{\includegraphics[width=0.5\textwidth,clip=]{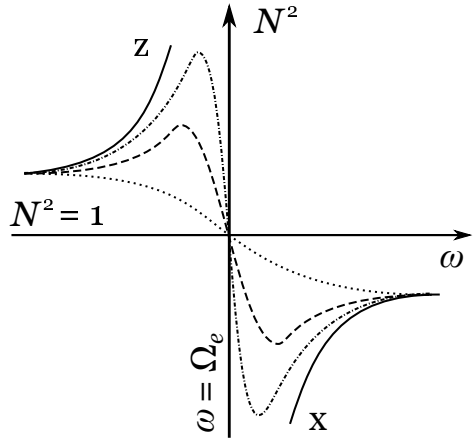}
              }
              \caption{Schematic of the washing-out of the stop band between the z-mode (solid curve labeled z) and x-mode (solid curve labeled x), which have a resonance, $N^2\to\infty$, at $\omega=\Omega_{\rm e}$ and a cutoff, $N^2=0$, at $\omega=\Omega_{\rm e}+\omega_{\rm p}^2/\Omega_{\rm e}$, with $-\infty<N^2<0$ inbetween. The dot-dash, dashed and dotted curves illustrate the effect of increasing electron temperature on the dispersion curve. The stop band is washed out when there is no region with $N^2<0$, with the line $N^2=0$ far below the region shown for $\omega_{\rm p}^2\ll\Omega_{\rm e}^2$. }
   \label{Fig:stopband}
   \end{figure}
   
\subsection{Washing-Out of the Stop Band}

ECME due to a ring or horseshoe distribution occurs at the relativistic gyrofrequency, which is at $\omega<\Omega_{\rm e}$ and so cannot be in the x-mode. The cold plasma dispersion relation for the extraordinary mode for perpendicular propagation in a plasma with $\omega_{\rm p}^2\ll\Omega_{\rm e}^2$ may be approximated by \citep{M13}
\be
N^2\approx\frac{\omega-\omega_x}{\omega-\Omega_{\rm e}},
\qquad
\omega_x\approx\Omega_{\rm e}+\omega_{\rm p}^2/\Omega_{\rm e}.
\label{z-xmode}
\ee
The mode separates into two branches: the x-mode applies for $\omega>\omega_x$ and the z-mode for $\omega<\Omega_{\rm e}$, with a stop band, $N^2<0$, in the range $\Omega_{\rm e}<\omega<\omega_x$. For perpendicular propagation, the generalization to include thermal effects involves intrinsically relativistic dispersion functions \citep{M13} and is not discussed here. For oblique propagation $k_\parallel\ne0$ the generalization involves only the familiar plasma dispersion function in a nonrelativistic thermal plasma. Semi-quantitatively, the right-hand side of the second of Equation (\ref{z-xmode}) is multiplied by the plasma dispersion function $\phi(y)$, with $y^2=(\omega-\Omega_{\rm e})^2/2k_\parallel^2V^2$, where $V$ is the thermal speed of the electrons. The cold plasma limit corresponds to $\phi(y)\to1$, $y^2\gg1$, and one has $\phi(y)\to2y^2$ for $y^2\ll1$. The effect of increasing $k_\parallel^2V^2$ is illustrated in Figure~\ref{Fig:stopband}. The stop band gets washed out as $k_\parallel^2V^2$ increases.  Vacuum dispersion is approximately valid for the cases indicated by the dashed and dotted curves. This occurs for \citep{M86}
\be
\frac{V}{c}|\cos\theta|\gtrsim\frac{\omega_{\rm p}^2}{\Omega_{\rm e}^2}.
\label{vdis}
\ee
When this condition is satisfied, the z-mode joins on continuously to the x-mode, and ring- or horseshoe-driven ECME results in escaping x-mode radiation.

\subsection{Back Reaction to ECME}

The back reaction to maser emission may be described using quasilinear theory. Numerical treatments for AKR \citep{P86,Petal02,KV12} show that the back reaction tends to drive the electrons to lower energy, tending to decrease the positive values of $\partial f(v)/\partial v$ and hence to suppress the instability. If suppression did occur, the back reaction would lead to substantial modification of the distribution function.  However, observation of a horseshoe distribution implies that any modification due to the back reaction is only minor, and hence that the maser is not operating near saturation. This allows one to put a limit on the intensity of the radiation: it must be much less than that corresponding to the saturation level. Assuming that the maser is driven by acceleration due to $E_\parallel$, the saturation level may be estimated by comparing the rate electrons lose energy to the waves with the rate their energy changes due to the acceleration. We outline such an estimate in Appendix~\ref{sect:saturation}.

We have no detailed data on whether or not the maser saturates in the solar case. However, there are several indirect arguments that suggest that saturation is unlikely. One argument is that the development of maser instabilities is likely to be highly intermittent. This is known to be the case for type-III radio bursts in the interplanetary medium, where the Langmuir waves produced by the instability are in localized, widely separated clumps, and this has been included in models for the radio emission \citep{MDC86}. The back reaction to the highly intermittent instability can be treated statistically \citep{MC89}, and the results suggest that the electron distribution is maintained at close to marginal stability by the many localized bursts of wave growth. A similar intermittency applies to $E_\parallel$: rather than $E_\parallel$ being slowly varying in space, it is highly localized, in regions of order a Debye length \citep{Eetal98}. One might expect that localized bursts of wave growth are associated with the localized regions of acceleration. As in the case of type-III bursts, it is plausible that highly intermittent wave growth occurs due to the distribution being close to marginal stability, with the back reactions to the statistically large number of localized bursts of growth maintaining this marginally stable state.

\section{Discussion}
\label{sect:discussion}

We concentrate the discussion here around the two questions posed in Section~\ref{s:introduction}. One concerns the likelihood of a horseshoe distribution developing in the acceleration of the electrons that produce HXRBs. The other concerns the relevance of horseshoe-driver ECME to solar spike bursts. 

\subsection{Do the Precipitating Electrons in HXRBs Have a Horseshoe Distribution?}

Our working hypothesis is that the acceleration of the electrons that emit HXRBs and generate type-III radio bursts in a solar flare is analogous to the acceleration of the auroral (inverted-V) electrons that generate AKR. This includes the assumption that energy is transported Alfv\'enically \citep{FH08,H12} between a generator region near the apex of the flux tube and an acceleration region near the chromosphere, where the electrons are accelerated by a parallel electric field \citep{MW13,MW14}.  The answer to this question is affirmative if the processes that lead to a horseshoe distribution for the electrons that generate  AKR also apply to the electrons that generate HXRBs.

In Section~\ref{sect:horseshoe} we formalize existing qualitative arguments for the formation of a horseshoe distribution in terms of three assumption in a conventional 1D model for the self-consistent parallel potential and current \citep{K73,W77,CS78}. Assumption (i) is that electrons are injected at the apex of a flux tube and accelerated towards a footpoint by a parallel electric field. Where these ``injected'' electrons ultimately come from is uncertain in both the magnetospheric and coronal applications. We favor upward acceleration from the chromosphere at the conjugate footpoint, as suggested by \citet{EH95}, but we do not discuss this point further in the present paper.  Our assumption (ii) is that the magnetic moment [$\mu$] is conserved. This conservation law breaks down in the presence of pitch-angle scattering, which is assumed to be absent. Collisions (Coulomb scattering), which are negligible in the auroral application, can be of marginal significance for  energetic electrons in a solar flux tube, and could lead to some pitch-angle scattering. The collision frequency for an electron with speed $v$ is proportional to $n_{\rm e}/(v/V_{\rm e})^3$, where $n_{\rm e}$ and $V_{\rm e}$ are the density and thermal speed of thermal electrons. The acceleration tends to evacuate the flux tube of thermal plasma, resulting in the observed density cavity in the auroral case \citep{BC79,Aetal15}. Assuming that analogous processes operate in the solar case, pitch-angle scattering due to collisions should be negligible.  Assumption (iii) is that electrons are efficiently scattered in the injection/generator region so that their distribution remains isotropic there. This condition is required for conservation of ${\cal E}$ and $\mu$ in the 1D model to imply a horseshoe distribution: the injected electrons must be isotropic. In the model, the electron distribution includes both downgoing electrons, which  remain isotropically distributed, and upgoing electrons, which are anisotropic due to a loss cone. If the upgoing electrons are not isotropized when they return to the injection region then the same distribution would apply in both hemispheres, and there would be a two-sided loss cone. A one-sided loss cone, which is observed for the electrons that generate AKR and that we assume also applies to the electrons in HXRBs, requires that on returning to the injection region, the electrons must effectively lose all memory of which hemisphere they came from. Pitch-angle scattering must be strong enough to fill the loss cone before the electrons escape the injection region again. Possible mechanisms that could lead to such efficient pitch-angle scattering confined to the generator region need to be discussed critically, but we do not do so here. With these three assumptions, a horseshoe distribution is to be expected. 

An interpretation of the processes that lead to a horseshoe distribution is as follows. Alfv\'enic energy transport involves oppositely directed, field-aligned currents on neighboring field lines, with these currents closing across the field lines in the generator region and in the chromosphere. The acceleration occurs on the field lines with upward current, carried by electrons that originate from the generator region. The only electrons that contribute to this current are those that do not mirror, that is those in the loss cone at the injection region. There is a limited supply of electrons in the injection region. The supply problem is further exacerbated by the field-aligned current between the apex and the footpoint being carried only by electrons in the loss cone at the apex. For $\alpha_{\rm L}\ll1$ only a fraction of order $\alpha_{\rm L}$ of this limited supply of electrons contribute to the current. The driving of the system requires a much larger current than can be supplied by these electrons alone. Our interpretation of how this supply problem is overcome has two ingredients. First, in a given flux tube (with a upward current) a parallel potential develops and accelerates electrons at just the rate needed to keep the field-aligned current constant, specifically to give $J(s)\propto B(s)$, except near the regions where the current closes across field lines. This effectively broadens the loss cone, towards $\pi/2$, maintains the distribution of downward propagating electrons nearly isotropic, and maximizes the fraction of electrons that are current-carriers. (The return current on neighboring field lines has no such restriction, due to the ample supply of chromospheric electrons.) Second, multiple current loops are  set up, in what we called a picket-fence model \citep{MW13}.

The energy distribution inferred for the precipitating electrons in a HXRB is a power law, but this does not invalidate the suggestion that these electrons locally have a horseshoe distribution. The horseshoe distribution observed in an inverted-V electron event is local, in the sense that the peak energy, corresponding to $v=v_0$, is a minimum at the edges of the relevant flux tube and a maximum in its center, resulting in the ``inverted-V'' energy spectrum observed by a spacecraft traversing the flux tube. When the electron distribution is integrated over the flux tube, the local ring-like energy spectrum is washed out. Although the spatial resolution of HXRBs is now good enough to lead to an estimate of the number density of the precipitating electrons \citep{Ketal11}, this applies to the spectrum integrated over a statistically large number of upward-current region in our picket-fence model \citep{H85,MW13}. It would be impossible to resolve a horseshoe distribution within an individual flux tube from the HXRB data. There is no inconsistency between the energy spectrum inferred from HXRB data and the assumption that locally the precipitating electrons have a horseshoe distribution.

\subsection{Does Horseshoe-Driven ECME Apply to Solar Spike Bursts?}

The foregoing discussion makes it plausible that a horseshoe distribution results from electron acceleration in a flaring flux tube. Another requirement for horseshoe-driven ECME to produce escaping x-mode radiation directly is that the plasma dispersion be essentially vacuum-like, specifically, that the stop band between the x~and z-modes be washed out. This requires a density cavity with effectively no cold plasma. Such a density cavity can form in the presence of downward acceleration by $E_\parallel$ when the supply of electrons is limited, so that all the available electrons are accelerated. Let us assume that such a coronal density cavity forms so that horseshoe-driven ECME operates in a solar flare, analogous to the generation of AKR in an inverted-V precipitation event. With this assumption, one would expect a close association between spike bursts and HXRBs: ECME should be a signature of precipitating electrons. However, not all HXRBs have associated spike bursts. For example, \citet{GAB91} found that while 95\,\% of spike bursts are associated with HXRBs, only 2\,\% of HXRBs have associated spike bursts. There is an important difference between the solar and magnetospheric cases that may explain the apparent absence of observable spike bursts in association with most HXRBs: gyromagnetic absorption.

Gyromagnetic absorption at the second harmonic is very strong in the solar corona, to the extent that it should prevent ECME from escaping \citep{MD82a}. All ECME must pass through the second-harmonic layer, and effectively all of it should be absorbed there. The relevant question becomes how any ECME gets through the second-harmonic layer. \citet{MWD89} considered various possible ways in which some radiation could escape, and concluded that even the most favorable way allowed only a small fraction to escape.  The fact that many HXRBs do not have observable spike bursts associated with them does not provide a strong argument against ECME occurring in association with all HXRBs: one expects most of the ECME to be absorbed at the second-harmonic layer. From this viewpoint, the relevant question is not why so few HXRBs have associated spike bursts, but rather what are the special conditions that allow any of the ECME to pass through this layer. 

The formation of a density cavity due to the acceleration by $E_\parallel$ suggests a new possibility for allowing a fraction of the ECME to escape through the second-harmonic layer. This possibility arises if the density cavity extends to above the second-harmonic absorption layer. This is the case when the flux tube in which the acceleration (and associated density depletion) occurs extends to a height where $B$ has decreased by a factor of two from its value at the emission point of the ECME. The radiation then passes through the second-harmonic layer in the low-density cavity. The gyromagnetic absorption coefficient is proportional to the density of thermal electrons, and hence would be anomalously weak in an anomalously low-density region. If the density in the cavity is orders of magnitude smaller inside the flux tube than outside it,  as is the case of AKR, then gyromagnetic absorption at the second harmonic would be unimportant. The fraction of the ECME that escapes would then be the fraction that is ducted along the low-density flux tube to above the second-harmonic layer. 

 \begin{figure}    
   \centerline{\includegraphics[width=0.5\textwidth,clip=]{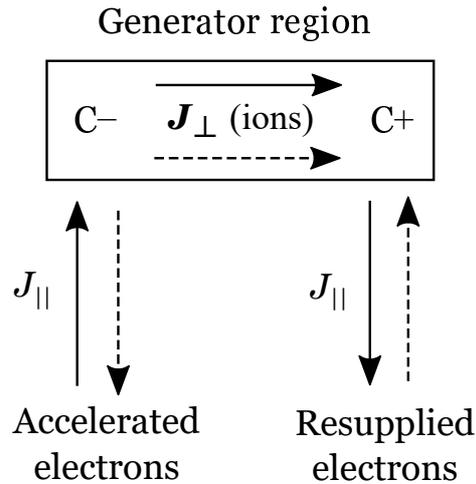}
              }
              \caption{Field-aligned (vertical) currents and cross-field (horizontal) currents join at corners C$\pm$. Both electrons and ions flow away from C$-$, leading to a density depletion, and towards C$+$, leading to a density enhancement.}
   \label{Fig:depletion}
   \end{figure}

\subsection{Do Coronal Cavities Exist?}

Whether or not a coronal density cavity forms in association with the acceleration by $E_\parallel$ of the electrons that produce HXRBs depends on the supply of electrons in the source region. If there is an ample supply of electrons available for acceleration, the acceleration has little effect on the plasma density. However, if the supply is limited, and the system is driven hard, as it is in a flare, then the acceleration of the electrons must create an anomalously low-density region. A detailed model for the supply of electrons  is needed to discuss this point further. Here we restrict our remarks to comments on relevant points already made in the literature.

The Alfv\'en-wave model for energy transport between a generator region and an acceleration region involves a closed current loop, with field-aligned currents carried by electrons, and cross-field closure at end points. \citet{EH95} assumed current closure in the chromosphere due to the Pedersen current, which is carried by ions. There are then four ``corners'' where field-aligned currents join onto cross-field currents. Let corners at which a cross-field current starts and ends be denoted C$\pm$, with the cross-field current flowing from C$-$ to C$+$. Both electrons and ions flow towards C$+$ and away from C$-$, implying that a density enhancement builds up at C$+$ and a density depletion at C$-$, as indicated in Figure~\ref{Fig:depletion}. \citet{EH95} pointed out that this kind of density cavity can be compensated by neutral particles, in the partially ionized chromosphere, flowing into the density cavity and being ionized there. The density depletion envisaged here arises from a similar effect in the generator region. As pointed out by \citet{Wetal02}, if the cross-field current in the generator region is carried by ions, as is the case for the inertial current in an Alfv\'en wave, the density depletion occurs where the upward current is redirected into a cross-field current in the generator region. The assumed source of the accelerated electrons, in our 1D model, is the generator region, and this corresponds to a C$-$. In this case there is no obvious source of plasma to compensate for the steady depletion at this C$-$ when a steady current flows around the closed loop. The resulting density cavity would then extend along the entire upward current path, from the injection/generator region to the footpoint in the chromosphere. This establishes that a density cavity does tend to form in association with the acceleration of precipitating electrons. It also emphasizes a related problem: the supply of electrons for continuing acceleration. We do not discuss these problems further in this paper.

\section{Conclusions}
\label{sect:conclusions}

Since the early 1980s, it has been widely accepted that spike bursts are solar counterparts of AKR, both due to ECME associated with precipitating energetic electrons. The loss-cone-driven version of ECME that was assumed (in the 1980s) to apply to AKR has since been replaced by a horseshoe-driven version of ECME, for which there is very strong observational evidence. In this paper we explore the implications that spike bursts are due to horseshoe-driven ECME. In Section~\ref{sect:horseshoe} we show that acceleration of electrons by $E_\parallel$ leads to the formation of a horseshoe distribution under conditions that plausibly apply in a solar flare. In Section~\ref{sect:ECME} we discuss ECME, and emphasize that the horseshoe-driven version requires an extremely low density of thermal plasma in the emission region. By analogy with the density cavity observed in the source region of AKR, we refer to this as a coronal density cavity. In Section~\ref{sect:discussion} we argue that both a horseshoe distribution and a coronal density cavity plausibly develop in association with HXRBs. 

We conclude that spike bursts are probably due to horseshoe-driven ECME. This requires that extremely low-density coronal cavities form in the solar corona in association with the acceleration of these electrons. We point out how such cavities form in an Alfv\'enic model involving the development of closed current loops between the generator region and the chromosphere. The existence of coronal cavities has implications for HXRBs if the cavities extend sufficiently deeply into the chromosphere, affecting the height at which the hard X~rays are generated. It also has implications on the escape of ECME through the second-harmonic gyromagnetic absorption layer if the cavities extend to above the second-harmonic absorption layer. Another requirement for the formation of a horseshoe distribution is that the electrons in the generator region be isotropic, and this requires effective pitch-angle scattering there. The existence of such scattering causes an individual electron to diffuse relatively slowly through the generator region, implying that an electron spends a long time in the loop-top region, with possible implications for loop-top hard X-ray sources.

\begin{acks}
We acknowledge support from an Australian Research Council Discovery Project grant. DBM acknowledges support from the International Space Science Institute, Bern, Switzerland, and discussions with members of the team on ``Magnetic Waves in Solar Flares.''  We thank an anonymous referee for helpful suggestions.

\smallskip
\noindent
{\bf Disclosure of Potential Conflicts of Interest} \quad The authors declare that they have no conflicts of interest.

\end{acks}

\appendix
\section{Derivation of the Horseshoe Distribution from the Boltzmann Equation}
\label{sect:Boltzmann}

An alternative derivation of the results for the 1D model presented in Section~\ref{sect:horseshoe} follows from the steady state collisionless Boltzmann equation with magnetic field convergence:
\begin{equation}\label{eq:ss_kinetic}
\vec{v}\cdot\grad f -\frac{e}{m}\vec{E}\cdot\grad_{\vec{v}}f=\left(\frac{\partial f}{\partial t}\right)_{\grad\vec{B}}.
\end{equation}
In the 1D case with $f=f(v,\cos\alpha ,s)$ and $\vec{E}=E\uvec{s}$ it follows that
\begin{equation}\label{eq:ss_1d_geometry}
\vec{v}\cdot\grad =v\cos\alpha\frac{\partial}{\partial s},\quad 
\vec{E}\cdot\grad_{\vec{v}}=E\left[\cos\alpha \frac{\partial}{\partial v}+\frac{1}{v}\sin^2\alpha \frac{\partial}{\partial \cos\alpha }\right].
\end{equation}
Using Equations~(\ref{eq:ss_1d_geometry}) together with the form for the magnetic field convergence term given by~\citet{McC92}, Equation~(\ref{eq:ss_kinetic}) becomes
\begin{equation}\label{eq:1d-ss-boltzmann}
v\cos\alpha\frac{\partial f}{\partial s}-\frac{e}{m}E\cos\alpha \frac{\partial f}{\partial v}-\frac{e}{m}\frac{E}{v}\sin^2\alpha \frac{\partial f}{\partial \cos\alpha }=\half v\sin^2\alpha \frac{{\rm d}\ln B}{{\rm d} s}\frac{\partial f}{\partial \cos\alpha }.
\end{equation}

Equation~(\ref{eq:1d-ss-boltzmann}) may be solved by characteristics. The distribution function is a constant: 
\begin{equation}\label{eq:fconst}
\frac{{\rm d}}{{\rm d}\xi}f\left[v(\xi),\cos\alpha  (\xi),s(\xi)\right]=0
\end{equation}
along characteristics defined by
\begin{eqnarray}
\frac{{\rm d}v}{{\rm d}\xi}&=&-\frac{e}{m}E\cos\alpha\label{eq:char1}\\
\frac{{\rm d}\cos\alpha }{{\rm d}\xi}&=&-\sin^2\alpha\left(\frac{e}{m}\frac{E}{v} +\half v \frac{{\rm d}\ln B}{{\rm d} s}\right),\label{eq:char2}\\
\frac{{\rm d}s}{{\rm d}\xi}&=&v\cos\alpha
\label{eq:char3}
\end{eqnarray}
where $\xi$ is a parameter along the path. Equation~(\ref{eq:fconst}) represents Liouville's theorem.

Writing $E=-{\rm d}\Phi/{\rm d}s$, and taking the ratio of Equations~(\ref{eq:char1}) and~(\ref{eq:char3}) gives an ODE which may be integrated to give
\begin{equation}\label{eq:energy_integral}
\half mv^2-e\Phi (s) ={\cal E},
\end{equation}
where ${\cal E}$ is the constant of integration. Similarly, taking the ratio of Equations~(\ref{eq:char2}) and~(\ref{eq:char3}) gives an ODE which may be integrated to give
\begin{equation}\label{eq:angle_integral1}
\frac{\sin^2\alpha}{\sin^2\alpha_{\rm inj}}=\frac{B(s)}{B_{\rm inj}}\frac{\cal E}{{\cal E}+e\Phi (s)},
\end{equation}
where we have used Equation~(\ref{eq:energy_integral}), and where $\alpha_{\rm inj}=\alpha(s_{\rm inj})$, $B_{\rm inj}=B(s_{\rm inj})$ and $\Phi (s_{\rm inj})=0$, following the notation of Section~\ref{sect:horseshoe}. Equation~(\ref{eq:angle_integral1}) may be rewritten
\begin{equation}\label{eq:angle_integral2}
\sin\alpha (s) = \sin\alpha_{\rm inj}\left[\frac{B(s)}{B_{\rm inj}}\frac{\cal E}{{\cal E}+e\Phi (s)}\right]^{1/2}.
\end{equation}
Equations~(\ref{eq:energy_integral}) and~(\ref{eq:angle_integral2}) are equivalent to Equations~(\ref{Hmu1}) in Section~\ref{sect:horseshoe}, and represent conservation of energy and magnetic moment respectively.

The results of Section~\ref{sect:horseshoe} follow from Equations~(\ref{eq:fconst}), (\ref{eq:energy_integral}), and~(\ref{eq:angle_integral2}). The argument is essentially the same but is restated here. 
The distribution in the flux tube consists of downgoing ($\frac{\pi}{2}<\alpha<\pi$) and upgoing electrons ($\alpha<\frac{\pi}{2}$). Downgoing electrons are lost at the footpoint if $\alpha_{\rm L\ast}= \alpha (s_{\ast})> \frac{\pi}{2}$. From Equation~(\ref{eq:angle_integral2}) this implies that downgoing electrons at the apex of the tube with an initial pitch angle in the range $\alpha_{\rm L\, inj}<\alpha\leq \pi$ are lost, where
\begin{equation}\label{eq:alphaL0}
\sin\alpha_{\rm L\, inj}=\left[\frac{B_{\rm inj}}{B_{\ast}}\frac{{\cal E}+e\Phi_0}{\cal E}\right]^{1/2}.
\end{equation}
From Equations~(\ref{eq:angle_integral2}) and~(\ref{eq:alphaL0}), electrons with pitch angle $\alpha_{\rm L\, inj}$ at the apex have a pitch angle
$\alpha_{\rm L}(s)$ at a point $s$ along the tube given by Equation~(\ref{alphaL}).
In the steady state upgoing electrons at the apex of the tube must exhibit a loss cone $0\leq \alpha\leq \pi-\alpha_{\rm L\, inj}$. Following Section~2.2.5, if we assume a Maxwellian form for the downgoing electrons at the apex, the complete distribution at $s=s_{\rm inj}$ must have the form
\be
f(v_{\rm inj},s_{\rm inj})=C\exp\left(-{\half mv_{\rm inj}^2/ T_{\rm inj}}\right)H[\alpha-\pi+\alpha_{\rm L\, inj}],
\label{eq:maxwellian_s0}
\ee
with $C=n(2\pi)^{-3/2}V_{\rm inj}^{-3}$. The distribution in the tube then follows from Equation~(\ref{eq:fconst}):
\begin{equation}
f(v,s)=f(v_{\rm inj},s_{\rm inj}).
\end{equation}
 Using Equation~(\ref{eq:energy_integral}) to express $v_{\rm inj}$ in terms of $v(s)$:
 \begin{equation}
 \half mv_{\rm inj}^2=\half mv^2-e\Phi (s),
 \end{equation}
and using Equation~(\ref{alphaL}) to transform the loss-cone angle at $s_{\rm inj}$ to that at position $s$ gives Equation~(\ref{mm3}), the horseshoe distribution:
\be
f(v,s)=C\exp\left(-{\half mv^2-e\Phi(s)\over T_{\rm inj}}\right)H[\alpha-\pi+\alpha_{\rm L}(s)].
\label{mm4}
\ee

\section{Saturation of Wave Growth}
\label{sect:saturation}

We use a quasilinear treatment of ECME, similar to that described by \citet{KV12}, both  to derive the growth rate, $\gamma_{\rm x}$, of x-mode waves,  and also to describe the back reaction on the electron distribution. For simplicity, we assume the vacuum limit ($N^2\to1$) and perpendicular propagation ($k_\parallel=0$). The resonance condition then reduces to $\omega=\Omega_{\rm e}(1-v^2/2c^2)$.

The growth rate is
\be
\gamma_{\rm x}={\pi^2e^2c^2\over m\varepsilon_0\Omega_{\rm e}}\langle\sin^2\alpha\rangle\left.\left(v^2{\partial f(v)\over\partial v}\right)\right|_{v=v_{\rm res}},
\qquad
v_{\rm res}=c\left[2\frac{\Omega_{\rm e}-\omega}{\Omega_{\rm e}}\right]^{1/2},
\label{gammax}
\ee
with $\langle\sin^2\alpha\rangle=(2-\cos^3\alpha_{\rm L})/3$ and $v^2\partial f(v)/\partial v=v(b-v^2/V^2)f(v)$, for the distribution function (\ref{mm1}) with $mV^2=T_{\rm inj}$.

In this case, the distribution function may be approximated by an isotropic diffusion in velocity space, with an evolution described by:
\be
\frac{\partial f(v)}{\partial t}={1\over v^2}{\partial\over\partial v}
\left(v^2D(v){\partial f(v)\over\partial v}\right).
\label{QL1}
\ee
Assuming x-mode waves propagating perpendicular to the field lines with an energy density $W_{\rm x}(\Delta\omega)$ per unit frequency, with $\Delta\omega=\Omega_{\rm e}-\omega$, we find
\be
D(v)={\pi e^2\sin^2\alpha\over4m^2\varepsilon_0}W_{\rm x}(\Delta\omega_{\rm res}),
\qquad
\Delta\omega_{\rm res}=\Omega_{\rm e}{v^2\over2c^2}.
\label{QL2}
\ee
The diffusion leads to an increase in energy spread that includes a systematic change in energy that may be described by
\be
\left\langle\frac{\rm d}{{\rm d}t}\half mv^2\right\rangle={m\over v^2}{\partial\over\partial v}[v^3D(v)].
\label{QL3}
\ee
The derivative in Equation (\ref{QL3}) changes sign as a function of $v$ due to $W_{\rm x}(\Delta\omega)$ having a maximum as a function of $\Delta\omega$. Higher-energy electrons resonate with lower-frequency waves and lose energy to the wave, and lower-energy electrons resonate with higher-frequency waves and gain energy from them; there is a net loss of energy to the waves because there are more higher-energy than lower-energy electrons. Saturation of the maser becomes significant when these changes become important in modifying the (assumed horseshoe) distribution.

A limit on the level of the waves for the maser to remain unsaturated follows by assuming that the rate of change of energy due to the quasilinear diffusion is small in comparison with that due to the acceleration. This corresponds to 
\be
\left|\left\langle\frac{\rm d}{{\rm d}t}\half mv^2\right\rangle\right|\ll e|E_\parallel|v.
\label{QL4}
\ee
Further discussion of this point requires a more detailed model that takes into account the expected intermittency of the maser.

%
%
 \bibliographystyle{spr-mp-sola}
 \bibliography{ECMERefs}  
%
%

\end{article} 
\end{document}